\documentclass{aastex62}

\graphicspath{{./}{figures/}}

\usepackage{graphicx}

\usepackage{xcolor}

\usepackage{appendix}

\usepackage{chngcntr}

\accepted{March 6, 2020}
\submitjournal{The Astrophysical Journal}

\received{February 26,2020}

\shorttitle{Kepler Ejecta Kinematics Study}
\shortauthors{Millard et al.}
\begin{document}

\title{An Ejecta Kinematics Study of Kepler's Supernova Remnant with High-Resolution {\it Chandra} HETG Spectroscopy
}

\author{Matthew J. Millard}
\affil{Box 19059, Department of Physics, University of Texas at Arlington, Arlington, TX 76019}

\author{Jayant Bhalerao}
\affiliation{Box 19059, Department of Physics, University of Texas at Arlington, Arlington, TX 76019}

\author{Sangwook Park}
\affiliation{Box 19059, Department of Physics, University of Texas at Arlington, Arlington, TX 76019}

\author{Toshiki Sato}
\affil{RIKEN, 2-1 Hirosawa, Wako, Saitama 351-0198, Japan}
\affil{NASA, Goddard Space Flight Center, 8800 Greenbelt Road, Greenbelt, MD 20771, USA}
\affil{Department of Physics, University of Maryland Baltimore County, 1000 Hilltop Circle, Baltimore, MD 21250, USA}

\author{John P. Hughes}
\affiliation{Department of Physics and Astronomy, Rutgers University, 136 Frelinghuysen Road, Piscataway, NJ 08854-8019, USA}
%\affiliation{Center for Computational Astrophysics, Flatiron Institute, 162 Fifth Avenue, New York, NY 10010, USA}

\author{Patrick Slane}
\affiliation{Harvard-Smithsonian Center for Astrophysics, 60 Garden
Street, Cambridge, MA 02138, USA}

\author{Daniel Patnaude}
\affiliation{Smithsonian Astrophysical Observatory, Cambridge, MA 02138, USA}

\author{David Burrows}
\affiliation{Dept. of Astronomy \& Astrophysics, Penn State University,
University Park, PA 16802 USA}

\author{Carles Badenes}
\affiliation{Department of Physics and Astronomy and Pittsburgh Particle
Physics, Astrophysics and Cosmology Center (PITT PACC),
University of Pittsburgh, 3941 O’Hara Street, Pittsburgh, PA
15260, USA}
\affiliation{Institut de Ci\`{e}ncies del Cosmos (ICCUB), Universitat de
Barcelona (IEEC-UB), Mart\'{i} Franqu\'{e}s 1, E08028 Barcelona,
Spain}



\begin{abstract}

We report our measurements of the bulk radial velocity from a sample of small, metal-rich ejecta knots in Kepler's Supernova Remnant (SNR). We measure the Doppler shift of the He-like Si K$\alpha$ line center energy in the spectra of these knots based on our {\it Chandra} High-Energy Transmission Grating Spectrometer (HETGS) observation to estimate their radial velocities.  We estimate high radial velocities of up to $\sim$ 8,000 km s\textsuperscript{-1} for some of these ejecta knots. We also measure proper motions for our sample based on the archival {\it Chandra} Advanced CCD Imaging Spectrometer (ACIS) data taken in 2000, 2006, and 2014. Our measured radial velocities and proper motions indicate that some of these ejecta knots are almost freely-expanding after $\sim$ 400 years since the explosion. The fastest moving knots show proper motions up to $\sim$ 0.2 arcseconds per year. Assuming that these high velocity ejecta knots are traveling ahead of the forward shock of the SNR, we estimate the distance to Kepler's SNR {\it d} $\sim$ 4.4 to 7.5 kpc. We find that the ejecta knots in our sample have an average space velocity of $ v\textsubscript{s} \sim$ 4,600 km s\textsuperscript{-1} (at a distance of 6 kpc). We note that 8 out of the 15 ejecta knots from our sample show a statistically significant (at the 90\% confidence level) redshifted spectrum, compared to only two with a blueshifted spectrum. This may suggest an asymmetry in the ejecta distribution in Kepler's SNR along the line of sight, however a larger sample size is required to confirm this result.

%If these ejecta knots are  significantly decelerated along the lines of sight by the reverse shock, we estimate {\it d} $\lesssim$ 11.2 kpc to Kepler's SNR.

\end{abstract}

%%\keywords{editorials, notices --- 
%%miscellaneous --- catalogs --- surveys}


\section{Introduction} \label{sec:intro}

	Type Ia supernova explosions are most likely the result of the unbinding of a white dwarf which has accreted enough mass from a companion, either through a merger or matter stream  \citep{1984ApJS...54..335I}, to burn carbon and oxygen  \citep{1960ApJ...132..565H}, resulting in a runaway thermonuclear explosion.  The evolution of Type Ia SNRs may be modelled assuming a uniform interstellar medium (ISM) interaction\citep{2007ApJ...662..472B,2018ApJ...865..151M}. However, asymmetries in ejecta distributions have been seen in some Type Ia SNRs (e.g., \citet{2013ApJ...771...56U,2014ApJ...792L..20P}), indicating that the explosion environment is likely more complex. The explosion itself might not have been spherically symmetric (e.g., \citet{2009Natur.460..869K,2010Natur.466...82M,2011MNRAS.413.3075M}), and the initial non-uniformity in the SN ejecta may be caused by such an explosion asymmetry. If the white dwarf is interacting with a non-degenerate companion star, the disk that would likely form around the accreting white dwarf may produce a wind which could strip material from the companion, creating an anisotropic circumstellar medium (CSM)  (e.g., \citet{2008ApJ...679.1390H}) surrounding the progenitor system.  Such a modified medium could contain regions of varying density, which may slow down some of the ejecta from the SN explosion, while leaving other parts of the ejecta gas unaffected.

%Recent theoretical (e.g., \citet{2009Natur.460..869K}) and observational evidence (e.g., \citet{2010Natur.466...82M,2011MNRAS.413.3075M}) has shown that Type Ia explosions may be asymmetrical events. An explosion asymmetry would in turn lead to an asymmetrical ejecta distribution in the resulting supernova remnant (SNR).  Further asymmetries in the ejecta distribution within an SNR may be due to interactions with a nonuniform medium into which it is expanding.

 A well-known case where a Type Ia SNR is interacting with CSM is the remnant of Supernova (SN) 1604, or Kepler's SNR (Kepler, hereafter),  the most recent Galactic historical supernova.  As a young, ejecta-dominated remnant of a luminous (assuming a distance $>$ 7 kpc) Type Ia SN \citep{2012ApJ...756....6P} from a metal-rich progenitor \citep{2013ApJ...767L..10P}, it provides an excellent opportunity to study the nature of a Type Ia progenitor and its explosion in the presence of CSM material \citep{2013ApJ...764...63B} and nitrogen-rich gas \citep{1982A&A...112..215D,1991ApJ...366..484B,2015ApJ...808...49K}. Strong silicate dust features observed in the infrared spectra of the remnant are indicative of the wind from an oxygen-rich asymptotic giant branch (AGB) star \citep{2012ApJ...755....3W}. The distance to Kepler's SNR is uncertain; recent estimates put the distance from 3.9 kpc \citep{2005AdSpR..35.1027S} to $\ga{}$7 kpc \citep{2012ApJ...756....6P,2012A&A...537A.139C}.

In X-rays, Kepler appears as mostly circular with an angular diameter of $\sim$ 3.6\arcmin{}, however it does have curious morphological features. For example, there are two notable protrusions located in the east and west portion of the SNR, often referred to as ``Ears" \citep{2013MNRAS.435..320T}  (a similar case is G299.2-2.9 \citep{2014ApJ...792L..20P}). Kepler also shows emission features from shocked CSM, one located across the center of the remnant and another which stretches across the northern rim \citep{2013ApJ...764...63B}. \citet{2013ApJ...767L..10P} found a higher Ni to Fe K line flux ratio in the northern half than in the southern half of Kepler, but were not able to distinguish the origin for the differential Ni/Fe flux ratio (shock interactions with different CSM densities between the north and south versus an intrinsically different ejecta distribution between the north and south). \citet{2008ApJ...689..225K} found that the northern half was expanding more slowly than the southern half, suggesting an uneven ejecta distribution between the northern and southern shells, although they attributed the difference to interaction with a dense CSM in the north.

Measuring the Doppler shifts in the emission lines from the X-ray-emtting ejecta knots projected over the face of the SNR, and thus their bulk motion line-of-sight (``radial'' hereafter) velocities ($v_{r}$) is useful to reveal the 3-D structure of the clumpy ejecta gas. The velocity measurements of these knots may help to reveal the ejecta properties immediately after the explosion, as well as their interaction with the circumstellar medium, which was formed by the progenitor system's mass loss history.
  Recently, \citet{2017ApJ...845..167S} reported measurements of radial velocity for several compact X-ray-bright knots in Kepler's SNR using archival {\it Chandra} ACIS data. They measured high radial velocities of up to $\sim 10^{4}$ km s\textsuperscript{-1} and nearly free-expansion rates for some knots. 
%Here, we present the results of our study on the 3-D structure of Kepler, based on the high resolution X-ray spectroscopy with our {\it Chandra} HETGS observation.

 Here, we present the results of our study on 3-D velocity measurements of a sample of 17 small, bright regions in Kepler, based on high resolution X-ray spectroscopy from our {\it Chandra} HETGS observation. In Section \ref{sec:obs}, we present the observations we used for our analysis.  In Section \ref{sec:data}, we show our analysis techniques and results.  In Section \ref{sec:discuss}, we estimate the distance to Kepler and discuss its ejecta distribution based on our results, and in Section \ref{sec:conclusions} we summarize our findings.

\section{Observations} \label{sec:obs}

We performed our {\it Chandra} HETGS observation of Kepler using the ACIS-S array from 2016 July 20 to 2016 July 23. The aim point was set at RA(J2000) = 17\textsuperscript{h}30\textsuperscript{m}41\textsuperscript{s}.3, Dec(J2000) = -21$^{\circ}$29\arcmin 28\arcsec.9, roughly towards the geometric center of the SNR. The observation was composed of a single ObsID, 17901.  We processed the raw event files using {\it Chandra} Interactive Analysis of Observations (CIAO) \citep{2006SPIE.6270E..60F} version 4.10 and the {\it Chandra} Calibration Database (CALDB) version 4.7.8 to create a new level=2 event file using the CIAO command, {\tt chandra\_repro}. Next, we removed time intervals of background flaring using the {\it Chandra} Imaging and Plotting System (ChIPS) command,  {\tt lc\_sigma\_clip},  which left us with a total effective exposure of 147.6 ks.  We then extracted the 1st-order dispersed spectra from a number of small regions across the SNR (Section \ref{subsec:radv}) using the TGCat scripts \citep{2011AJ....141..129H} {\tt tg\_create\_mask}, {\tt tg\_resolve\_events}, and {\tt tgextract}, and also created appropriate detector response files. The TGCat commands (in the order mentioned) first create a FITS region file which specifies a region position, shape, size, and orientation in sky pixel-plane coordinates\footnote{http://cxc.harvard.edu/ciao/ahelp/tg\_create\_mask.html}. Next, event positions are compared with the 3-D locations at which dispersed photons can appear, given the grating equation and zero order position, and TGCat assigns them a wavelength and an order, and outputs these data into a grating events file\footnote{http://cxc.harvard.edu/ciao/ahelp/tg\_resolve\_events.html}. Finally, the grating events file is filtered and binned into a one-dimensional counts spectrum for each grating part, order, and source\footnote{http://cxc.harvard.edu/ciao/ahelp/tg\_extract.html}. In addition to our new HETGS data, we also used the archival ACIS data of Kepler as supplementary data (listed in Table \ref{table:obs}). For spectral fitting  purposes (Section \ref{subsec:specmod}), we reprocessed the six ObsIDs from the 2006 archival ACIS-S3 data by following standard data reduction procedures with CIAO versions 4.8 to 4.8.2 and CALDB version 4.7.2, which resulted in a total effective exposure of $\sim$ 733 ks. To make our proper motion measurements, we used the 2000, 2006, and 2014 archival {\it Chandra} ACIS data, as previously processed and prepared in  \citet{2017ApJ...845..167S}.

\section{Data Analysis and Results} \label{sec:data}

\subsection{Utility of HETGS for Extended Sources} \label{subsec:hetgs}

\begin{figure}
\plotone{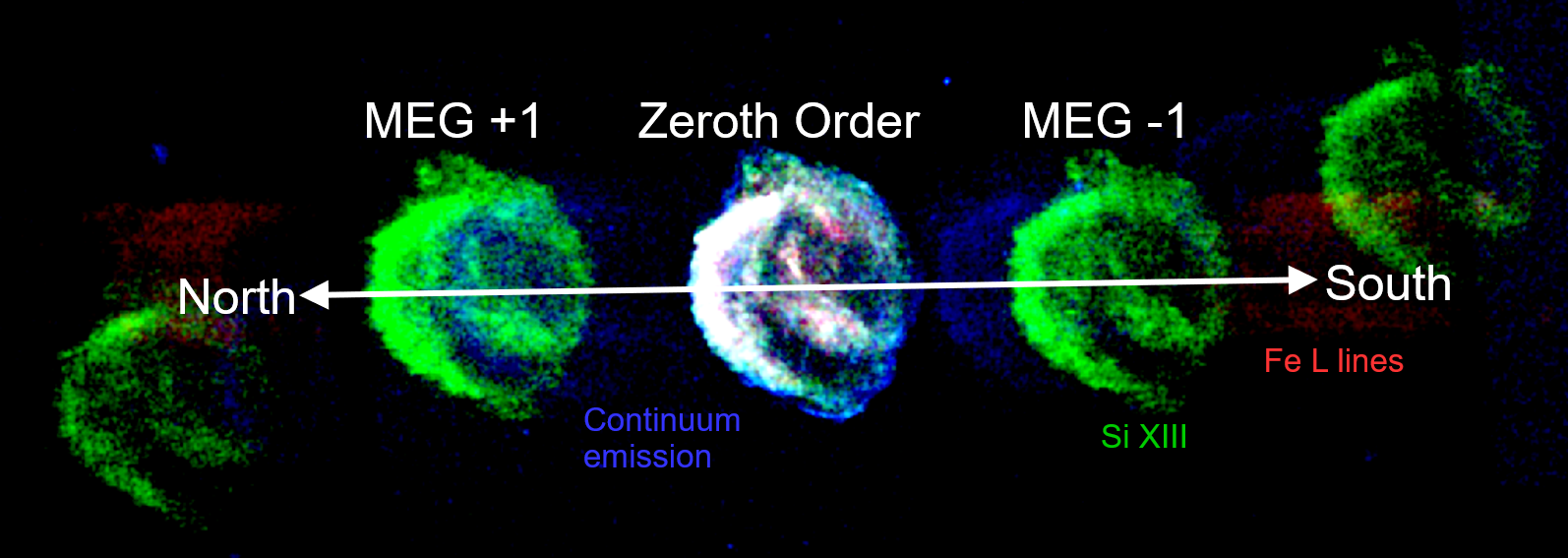}
\caption{{\it Chandra} HETG 3-color image of Kepler. Red: 0.7-1.2 keV, Green: 1.7-2.0 keV and Blue: 2.0-8.0 keV.  The Fe L complex and continuum emission appear smeared across the ACIS-S chips, the former because it consists of many emission lines, and the latter because it lacks individual emission lines.  The Si XIII (He-like Si K$\alpha$) emission is more focused on the detector, because it consists only of three closely spaced lines at $\sim$ 1.865 keV.
\label{fig:3color}}
\end{figure}

Due to its dispersed nature, the {\it Chandra} HETGS (the 1st-order) is best suited to measure the spectra of isolated, point-like sources.  The utility of the HETG spectrum is affected when the source is extended and/or surrounded by complex background emission features.  Our study of Kepler is typical of such a case; the SNR comprises many small, discrete extended sources projected against its own complex diffuse emission. The HETG-dispersed image of Kepler is shown in Figure \ref{fig:3color}.  Our goal is to measure the atomic line center energies in the X-ray emission spectrum for small individual emission features within the SNR. For this type of measurement, the utility of HETG data have been successfully demonstrated by previous authors in the cases of Cassiopeia A (Cas A) \citep{2006ApJ...651..250L} and G292.0+1.8 (G292) \citep{2015ApJ...800...65B}. He-like Si K$\alpha$ lines were used for Cas A, while He- and H-like Ne, Mg, and Si K$\alpha$ lines were used for G292.  In the integrated spectrum of Kepler, the Fe L and K, and He-like Si and S K$\alpha$ lines are prominent. However, the Fe K line is faint in the spectra of individual small knots, and thus, not useful for our study. Additionally, the Fe L lines are a complex composed of several closely spaced emission lines, which makes it difficult to identify them for Doppler shift measurements, whereas the He-like Si K$\alpha$ and S He-like K$\alpha$ lines each may easily be represented by a simple trio of emission lines.  Overall, ejecta knots in Kepler are fainter than those in Cas A and G292. Thus, the count statistics for most knots only allow us to use the brightest line, He-like Si K$\alpha{}$. In general, we found that at least $\sim$ 100 counts for the He-like Si K$\alpha$ line emission features in the 1.75 - 1.96 keV band of the 1st-order MEG (Medium Energy Grating) spectrum of each individual target source are required to make a reliable Doppler shift measurement.

%% The "ht!" tells LaTeX to put the figure "here" first, at the "top" next
%% and to override the normal way of calculating a float position

\startlongtable
\begin{deluxetable*}{ccC}
\tablecaption{Archival {\it Chandra} ACIS Observations \label{tab:obs}}
\tablecolumns{3}
\tablenum{1}
\tablewidth{0pt}
\tablehead{
\colhead{Observation ID} &
\colhead{Start Date} &
\colhead{Exposure Time (ks) }}
\startdata
116	 & 	2000-06-30	 & 	48.8	\\
4650	 & 	2004-10-26	 & 	46.2	\\
6714	 & 	2006-04-27	 & 	157.8	\\
6715	 & 	2006-08-03	 & 	159.1	\\
6716	 & 	2006-05-05	 & 	158.0	\\
6717	 & 	2006-07-13	 & 	106.8	\\
6718	 & 	2006-07-21	 & 	107.8	\\
7366	 & 	2006-07-16	 & 	51.5	\\
16004	 & 	2014-05-13	 & 	102.7	\\
16614	 & 	2014-05-16	 & 	36.4	\\
\enddata
\label{table:obs}
\end{deluxetable*}

\begin{figure}
\epsscale{0.9}
\plotone{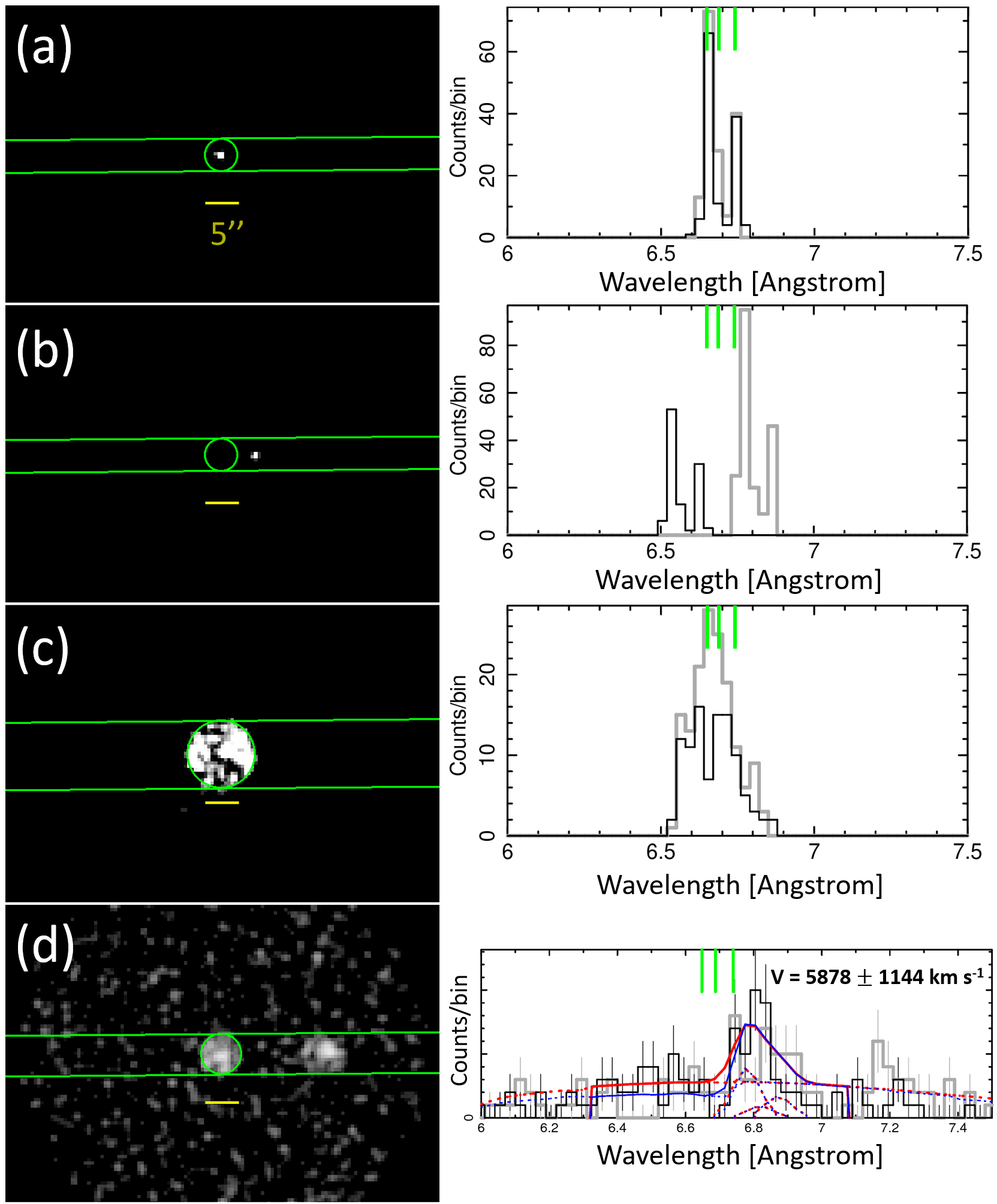}
\caption{The left column shows images from our MARX simulations (assuming a {\it Chandra} HETGS + ACIS-S configuration). The green circle and horizontal lines show the target region, and dispersion direction, respectively. The yellow scale bar in each image is 5\arcsec{} across.  The right column shows the extracted 1\textsuperscript{st} order MEG spectrum. The black line and gray line are the plus and minus order spectrum, respectively. The rest energies of the He-like Si K$\alpha$ line used here are 6.648 \AA{}, 6.688 \AA{}, and 6.740 \AA{}, denoted by green vertical lines in panel (d). Panel (a) shows the spectrum of a point source, with the zeroth order point centered on it. The plus and minus order spectra are aligned at the rest energy of the line trio when the source is located at the zeroth order point. (b) exhibits the effect of shifting the zeroth order point by 5\arcsec{} along the dispersion direction.  The plus and minus orders move away from the line center energy. (c) shows an extended source (10\arcsec{}), which broadens the resulting peaks in the spectra. (d) contains a complex source configuration. The target source has an angular size of 6\arcsec{}, and assumed radial velocity of $v_{r}$ = $+$ 6,000 km s\textsuperscript{-1}. The nearby source has an angular size of the 6\arcsec{} and an assumed radial velocity of $v_{r}$ = $+$ 9,000 km s\textsuperscript{-1}. The assumed angular offset of the nearby source is 15\arcsec{}. Despite the proximity of the two sources, the correct Doppler shift was measured, $v_{r}$ = 5878 $\pm$ 1144 km s\textsuperscript{-1}, in part due to the appropriate choice of osort value (0.05 in this case). The red line and blue lines are the best-fit models for the +1 and -1 order spectra, respectively.
\label{fig:marx}}
\end{figure}

Distinguishing the He-like Si K$\alpha$ lines from the target emission knot from those of the surroundings is essential to correctly measure the line center energies of He-like Si K$\alpha$ lines in the spectrum of small individual knots in Kepler. To quantitatively assess the contamination in the He-like Si K$\alpha$ line profiles of the target source from the nearby emission features, as well as due to the target source extent, we performed ray-trace simulations of {\it Chandra} observations using the Model of AXAF Response to X-rays (MARX) package \citep{2012SPIE.8443E..1AD}. Initially, we assumed a point-like target source with an X-ray spectrum representing the rest energy emission lines of He-like Si K$\alpha$, at various distances from the zeroth order position.  Figures \ref{fig:marx}a and \ref{fig:marx}b show that the 1st-order spectral lines (He-like Si K$\alpha$) are shifted from the true line center energies as the source position is off-centered, corresponding to the {\it Chandra} HETGS wavelength scale of 0.0113 \AA/arcsec for HEG (High Energy Grating) and 0.0226 \AA/arcsec for MEG\footnote{http://cxc.harvard.edu/proposer/POG/html/chap8.html}. Using this relation, we may identify interfering emission lines in our source spectra originating from nearby sources. We also tested how the angular extent of the target sources affect our line center measurements. While larger source extents would increase the uncertainties in the line center energy measurements, we conclude that our radial velocity measurements would not be affected (within uncertainties) as long as the target source sizes are $\la$ 10\arcsec{} (Figures \ref{fig:marx}c and \ref{fig:marx}d).

Based on our test simulations, we also conclude that nearby discrete sources positioned $\sim$ 25\arcsec{} or farther off the target source position along the dispersion direction would not affect our measurements of the source spectral line center energies for radial velocities. For the cases where nearby sources are present (with angular extent similar to that of the target source) within $\sim$ 25\arcsec{} of the target source along the dispersion direction, the effects on the line center measurements for the target source may vary. We investigated numerous source configurations (both with our actual data of Kepler and extensive MARX simulations), and found that even if the nearby source positions are relatively close to the target position, we may avoid a significant contamination from the nearby emission by adjusting the criteria for the selection of the 1st-order photons of the target spectrum via the ``osort" parameters, {\it osort\_lo} and {\it osort\_hi}\footnote{See footnote 2}. During HETG spectrum extraction, only photons with measured wavelengths that meet the criteria, $osort\_lo < \lambda_{g}/\lambda_{CCD} \leq osort\_hi$ are included in the 1st-order spectrum, where  $\lambda_{CCD}$ is the ACIS-S CCD wavelength, and $\lambda_{g}$ is the gratings wavelength. Because $\lambda_{CCD}$ and $\lambda_{g}$ values of nearby sources become more divergent the farther they are located from the target position, photons from those nearby sources are less likely to be included in the extracted spectrum when small osort values are chosen.
Thus, we may still be able to measure the source line center energies despite the presence of nearby contaminating emission features. However, we find it unlikely that the emission lines from sources  located very near to each other ($\la 5 \arcsec{}$) along the dispersion direction, with similar brightness, would be properly distinguishable.

%% The "ht!" tells LaTeX to put the figure "here" first, at the "top" next
%% and to override the normal way of calculating a float position

\subsection{Radial Velocities} \label{subsec:radv}

\begin{figure}
\plotone{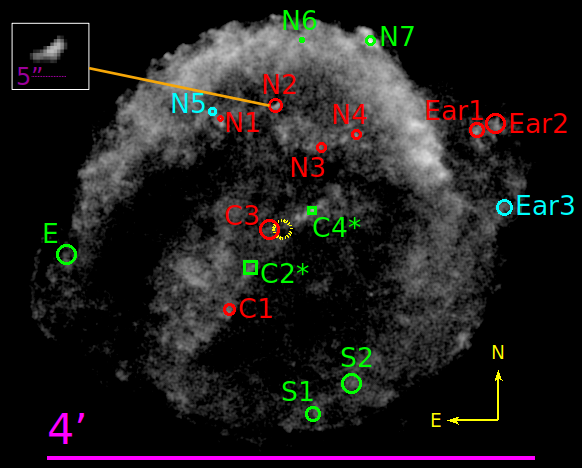}
\caption{ACIS-S gray-scale image of Kepler's SNR from the 2014 observation, filtered to the energy range 1.7 to 2.0 keV. Seventeen ejecta and CSM knots which we analyzed in this work are marked with circles.  CSM knots are marked with squares (also, their region names include `` * "). Otherwise, we identify all other knots to be metal-rich ejecta based on our spectral analysis of the archival ACIS data. Cyan and red markers indicate blue- and red-shifted features, respectively, while green represents statistically negligible {\it v\textsubscript{r}} at the 90\% confidence interval. The uncertainty in the kinematic center of the SNR estimated by \citet{2017ApJ...845..167S} is denoted by a dotted yellow circle. A zoomed-in image of knot N2 is shown in the upper left corner.
\label{fig:knotloc}}
\end{figure}

Based on archival {\it Chandra} ACIS data (Table \ref{table:obs}), we identified numerous small emission features which are bright in the 1.7 - 2.0 keV band, suggesting that they may be good candidate targets for He-like Si K$\alpha$ line center energy measurements using an HETG 1st-order spectrum. We selected 17 features (Figure \ref{fig:knotloc}), generally satisfying the criteria that we discussed in Sec \ref{subsec:hetgs}. 
To measure the {\it v\textsubscript{r}} of these X-ray emission features projected within the boundary of Kepler's SNR, we adopt a similar method to those pioneered by \citet{2006ApJ...651..250L} and \citet{2015ApJ...800...65B}, who analyzed HETG spectra of bright X-ray knots in SNRs Cas A and G292, respectively.    For each of these 17 individual features, we extracted the 1st-order spectrum from our {\it Chandra} HETGS observation.  

For each extracted region, the line center energies of the He-like Si K$\alpha$ lines, and two Si XII emission lines (see below), were measured by fitting six Gaussian curves to the spectrum - three for He-like Si K$\alpha$, two for the Si XII emission lines, and one for the background continuum using the Interactive Spectral Interpretation System (ISIS) software package \citep{2000ASPC..216..591H}.  The measured line center wavelengths were then compared with the rest values (6.648 \AA{} for resonance, 6.688 \AA{} for intercombination, 6.740 \AA{} for the forbidden line, and 6.717 \AA{} and 6.782 \AA{} for the Si XII lines, respectively \citep{1988CaJPh..66..586D}), to measure the Doppler shifts in these lines, and thus to estimate the corresponding {\it v\textsubscript{r}}.

The count statistics of our data do not allow us to directly measure the He-like Si (XIII) K$\alpha$ intercombination to resonance ({\it  i$/$r}) and forbidden to resonance ({\it f$/$r}) resonance line flux ratios. Thus, we use {\it  i$/$r} and {\it f$/$r} ratios which correspond to the values that we measured for each knot using archival ACIS data (Section \ref{subsec:specmod}). At the temperatures and ionization timescales that we measure for the knots in our sample, Si XII emission lines at $\sim$ 6.717 \AA{} and $\sim$ 6.782 \AA{} may also contribute significantly to the spectrum. Thus, we account for these lines in our {\it v\textsubscript{r}} fitting model. In the Appendix, we discuss the effects of varying line ratios on our $v_{r}$ estimates. In general, the uncertainty in line ratio values does not affect our conclusions.

\begin{figure*}[!htbp]
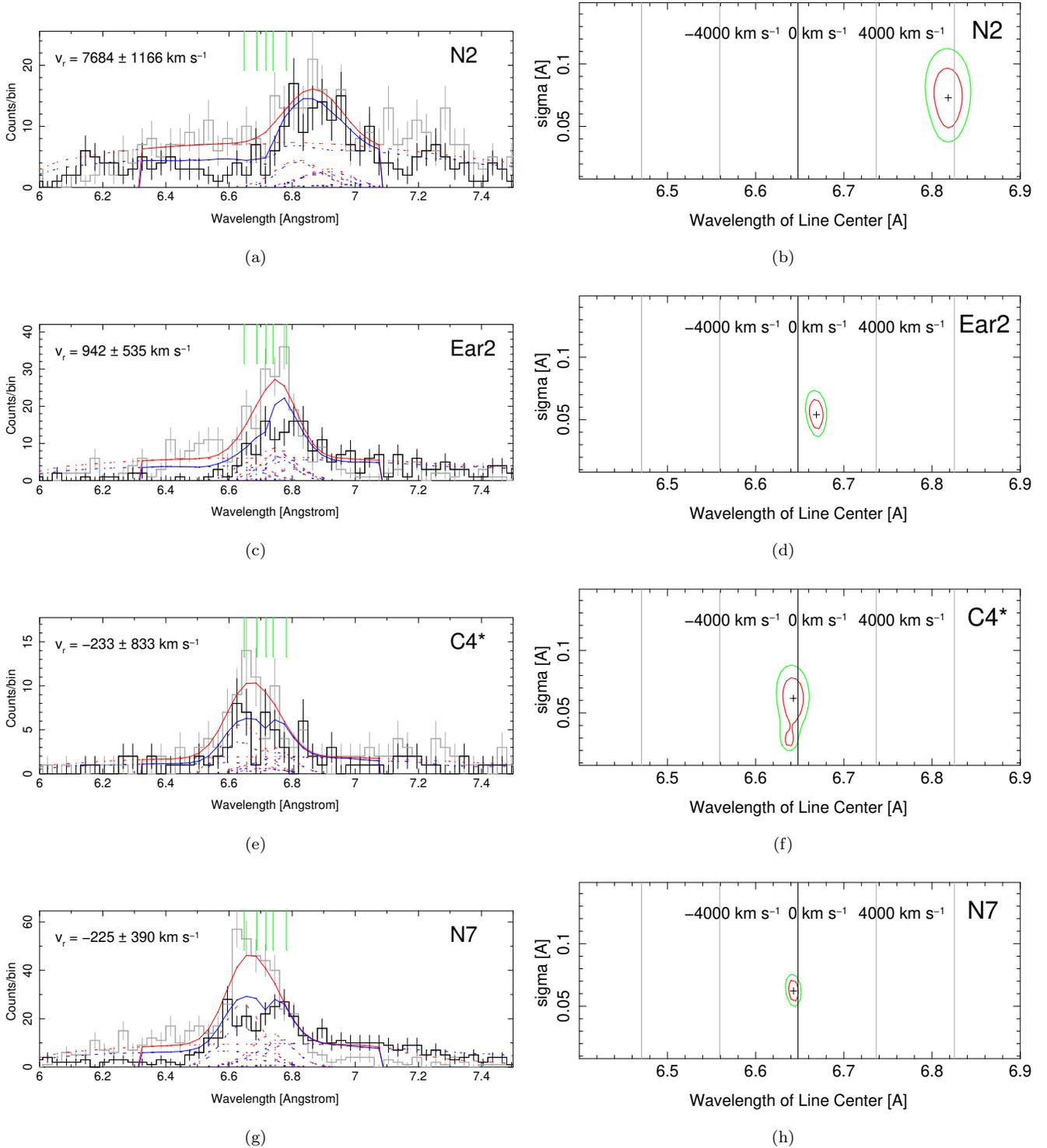

\gridline{\fig{N2_fitcheck2.eps}{0.49\textwidth}{(a)}
          %\fig{N2_contour_04_19_2019.png}{0.49\textwidth}{(b)}
          \fig{N2_contour.eps}{0.49\textwidth}{(b)}
          }
\gridline{\fig{Ear2_fitcheck2.eps}{0.49\textwidth}{(c)}
          \fig{Ear2_contour.eps}{0.49\textwidth}{(d)}
          }
\gridline{\fig{C4_fitcheck2.eps}{0.49\textwidth}{(e)}
          \fig{C4_contour.eps}{0.49\textwidth}{(f)}
          }
\gridline{\fig{N7_fitcheck2.eps}{0.49\textwidth}{(g)}
          \fig{N7_contour.eps}{0.49\textwidth}{(h)}
          }
\caption{Examples of our line center energy fits for small emission features in Kepler.  The left column is the HETG spectra overlaid with the best-fit model.  The straight green lines show the locations of the rest frame He-like Si K$\alpha$ line center wavelengths.  The dashed lines show individual Gaussian components of our best-fit model. The errors represent a 90$\%$ confidence interval. Gray: MEG +1 data,  Black: MEG -1 data, Red: MEG +1 model fit, Blue: MEG -1 model fit. The right column shows the confidence level contours for the best fit {\it v\textsubscript{r}} value. The red and green contours represent a 68$\%$ and 90$\%$ confidence interval, respectively. Panels (a-b) and (c-d) are from regions N2 and Ear2, respectively, showing clearly redshifted spectra. Panels (e-f) and (g-h) are from regions C4* and N7, respectively, showing negligible Doppler shift.} \label{fig:spectra}
\end{figure*}

Our results are summarized in Table \ref{table:all}, with spectra and best-fit models shown in Figure \ref{fig:spectra}. Errors represent a 90\% confidence interval unless otherwise noted. Figure \ref{fig:knotloc} shows the locations of blueshifted and redshifted regions, marked by cyan and red circles, respectively. Our measured {\it v\textsubscript{r}} for two CSM regions (regions C2* and C4*) are negligible even though they are projected near the SNR center. This low {\it v\textsubscript{r}} is perhaps as expected for the shocked CSM features regardless of  their projected distance from the SNR center, supporting the reliability of our {\it v\textsubscript{r}} measurements. Of the 15 ejecta knots for which we measured  {\it v\textsubscript{r}}, only two (N5 and Ear3) show a significantly blueshifted spectrum, while the other eight regional spectra are significantly redshifted.

We note that four ejecta knots in our sample (regions N2, N1, N3, and N7) were also studied in \citet{2017ApJ...845..167S}, who measured the {\it v\textsubscript{r}} of these ejecta knots based on {\it Chandra} ACIS data. For three of these common regions, N2, N1, and N3, we measure high  {\it v\textsubscript{r}} values ($v_r \sim$ 5,600 - 7,700 km s\textsuperscript{-1}). These regions are located in the northern shell of the SNR, approximately 1\arcmin{} from the kinematic center (R.A.(J2000) = 17\textsuperscript{h} 30\textsuperscript{m} 41\textsuperscript{s}.321 and Declination(J2000) = -21$^{\circ}$ 29\arcmin{} 30\arcsec.51 \citep{2017ApJ...845..167S}).
For all four of our common knots, we find general agreement between our measured values and those from \citet{2017ApJ...845..167S},  as shown in Table \ref{table:all}. This is an interesting result when we consider that the {\it v\textsubscript{r}} measurements based on the low-resolution ACIS spectrum are dominated by systematic uncertainties ($\sim$ 500 - 2,000 km s\textsuperscript{-1}) \citep{2017ApJ...840..112S,2017ApJ...845..167S}, while those using our high-resolution HETGS spectrum are mostly dominated by statistical uncertainties (due to the relatively lower throughput of the dispersed spectroscopy), yet our results for those four ejecta regions are consistent. Almost all other ejecta knots show significantly lower velocities of {\it v\textsubscript{r}} $\la$ 2,300 km s\textsuperscript{-1} (Table \ref{table:all}). It is notable that two ejecta knots (regions Ear1 and Ear2) projected within the western ``Ear" region show a significant {\it v\textsubscript{r}} ($\sim$ 2,200 and 900 km s\textsuperscript{-1}) even though they are projected far ($\sim$ 2\arcmin) from the center of the remnant beyond the main shell of the SNR.

\subsection{Identifying Metal-Rich Ejecta} \label{subsec:specmod}
  To identify the origin of small emission regions in our sample (metal-rich ejecta vs low-abundant CSM), we performed  spectral model fits for each individual regional spectrum based on the archival {\it Chandra} ACIS data with the deepest exposure (combining all ObsIDs taken in 2006, with a total exposure of 733 ks). We fitted the observed 0.3-7.0 keV band ACIS spectrum extracted from each region with an absorbed X-ray emission spectral model assuming optically-thin hot gas with non-equilibrium ionization ({\tt phabs*vpshock} \citep{2001ApJ...548..820B}) using the XSPEC software package \citep{1996ASPC..101...17A}. We estimated the background spectrum with small faint diffuse emission regions nearby each source region within the SNR. Then, we subtracted the background spectrum from the source regional spectrum before the spectral model fitting. We allowed the electron temperature($kT$, where $k$ is the Boltzmann constant and $T$ is electron temperature), ionization timescale ({\it n\textsubscript{e}t}: the electron density, {\it n\textsubscript{e}}, multiplied by the time since being shocked, {\it t}), redshift, normalization, and abundances of O, Ne, Mg, Si, S, Ar, and Fe to vary in the spectral model fitting. We fixed all other elemental abundances at solar values \citep{2000ApJ...542..914W}. We note that, although H and He are generally not expected to be abundant in the spectrum of ejecta-dominated emission features of Type Ia SNRs, Kepler is interacting with a significant amount of CSM. Thus, we leave the H and He abundances fixed at solar values in our model to account for possible CSM interaction throughout the SNR. In the ejecta-dominated knots, we also use the H and He continuum as an approximation for non-thermal power-law emission from the shock-accelerated electrons at the forward shock. We fixed the absorption column to {\it N}\textsubscript{H} = 5.4 $\times$ 10\textsuperscript{22} cm\textsuperscript{-2} \citep{2016ApJ...826...66F}.  We found significant residuals at $E$ $\sim$ 0.75 and $\sim$ 1.25 keV in the spectra of the ejecta knots. Similar features have been noticed in several other SNR studies of ejecta-dominated (particularly Fe-rich) spectra \citep{1998ApJ...497..833H,2015ApJ...808...49K,2017ApJ...845..167S, 2017ApJ...834..124Y}. To improve our spectral model fits, we added gaussian components at these energies to account for these emission line features of the ejecta knot spectra (adding these gaussian components does not significantly improve the spectral model fit of the CSM-dominated regions, and thus, were not included in those fits). We note that this implementation is not physically motivated, and is only intended as a statistical improvement in the spectral model fits. We confirm that excluding these gaussians does not affect our conclusions (distinguishing between ejecta-dominated and CSM-dominated regions, as discussed later). Our reduced chi-squared values for the best-fit models range from $\chi^{2}_{\nu} = $ 1.0 - 1.4.

%The addition of these gaussians does not affect our determination of which regions are ejecta-dominated or CSM-dominated.
 
\startlongtable
\begin{deluxetable*}{ccCcccccccc}
\tablecaption{Radial Velocity and Proper Motions of Small Emission Features in Kepler's SNR \label{tab:mathmode}}
\tablecolumns{12}
\tablenum{2}
\tablewidth{0pt}
\tabletypesize{\scriptsize}
\tablehead{
\colhead{Region\tablenotemark{$\dagger$}} &
\colhead{R.A.\tablenotemark{a}} &
\colhead{Dec\tablenotemark{a}} &
\colhead{{\it D}\tablenotemark{b}} &
\colhead{{\it v\textsubscript{r}}} &
\colhead{{\it v\textsubscript{r(SH)}}\tablenotemark{c}} &
\colhead{$\mu_{RA}$} & \colhead{$\mu_{Dec}$} & \colhead{$\mu_{Tot}$\tablenotemark{d}} & \colhead{$\eta$\tablenotemark{e}} &
\colhead{{\it v\textsubscript{s}}\tablenotemark{f}}  \\ 
\colhead{} & \colhead{(degree)} &
\colhead{(degree)} & \colhead{(arcmin)} & \colhead{(km s\textsuperscript{-1})} & \colhead{(km s\textsuperscript{-1})} & \colhead{(arcsec yr\textsuperscript{-1})} & \colhead{(arcsec yr\textsuperscript{-1})} & \colhead{(arcsec yr\textsuperscript{-1})} & & \colhead{(km s\textsuperscript{-1})}
}
\startdata
N2	 & 	262.67314	 & 	-21.474812	 & 1.02 & $	7684^{+1155}_{-1177}$ & $9110_{	-110}^{	+30}$ &	0.028	 $\pm	0.017	$  & 	0.137	 $\pm	0.024	$  & 	0.140	 $\pm	0.029	$  & $	0.94\pm $0.14 & $8656^{+1093}_{-1112}$	\\
N1	 & 	262.68120	 & 	-21.476634	 & 1.04 & $6019^{+1294}_{-1385}$ & $	8700	_{	-470	}^{	+650    	}$	& -0.065	 $\pm	0.016	$  & 	0.081	 $\pm	0.024	$  & 	0.104	 $\pm	0.028	$  & 	$0.68\pm $0.10  & $6707^{+1213}_{-1291}$		\\ 
N3	 & 	262.66648	 & 	-21.480553	 & 0.74  & $5550^{+2253}_{-2172}$ & $	5880	_{	-1750	}^{	+690    	}$	& 0.045	 $\pm	0.016	$  & 	0.078	 $\pm	0.024	$  & 	0.090	 $\pm	0.028	$  & 	$0.83\pm $0.16  & $	6113^{+2073}_{-2000}$		\\
C3	 & 	262.67403	 & 	-21.491796	 & 0.10 & $2281^{+1449}_{-1337}$ & 	& 0.019	 $\pm	0.018	$  &  -0.021	 $\pm	0.024	$  & 	0.028	 $\pm	0.03	$  & 	$1.86\pm $2.64  & $2416^{+1397}_{-1293}$		\\
N4	 & 	262.66124	 & 	-21.478912	 & 0.98 &$2252^{+1761}_{-1664}$ & 	& 0.061	 $\pm	0.017	$  & 	0.105	 $\pm	0.027	$  & 	0.121	 $\pm	0.032	$  & 	$0.85\pm $0.13  & $4115^{+1229}_{-1187}$		\\  
Ear1	 & 	262.64352	 & 	-21.478218	 & 1.79 & $2180^{+778}_{-752}$ & 	& 0.067	 $\pm	0.016	$  & 	0.059	 $\pm	0.024	$  & 	0.089	 $\pm	0.029	$  & 	$0.34\pm$ 0.03  & $3342^{+806}_{-795}$		\\
C1	 & 	262.67997	 & 	-21.502758	 & 0.79 & $1823^{+1408}_{-1458}$ & 	& -0.058	 $\pm	0.017	$  & 	-0.064	 $\pm	0.024	$  & 	0.086	 $\pm	0.029	$  & 	$0.75\pm$ 0.14  & $3052^{+1070}_{-1094}$		\\ 
Ear2	 & 	262.64087	 & 	-21.477378	 & 1.95 & $942^{+525}_{-546}$ & 	& 0.172	 $\pm	0.017	$  & 	0.104	 $\pm	0.024	$  & 	0.201	 $\pm	0.029	$  & 	$0.71\pm$ 0.05  & $5798^{+819}_{-819}$		\\ 
N6	 & 	262.66920	 & 	-21.465973	 & 1.56 & $	883^{+954}_{-933}$ & 	& 0.002	 $\pm	0.018	$  & 	0.141	 $\pm	0.024	$  & 	0.141	 $\pm	0.03	$  & 	$0.62\pm$ 0.06  & $	4109^{+859}_{-858}$		\\ 
E	 & 	262.70386	 & 	-21.495295	 & 1.78 & $531^{+1551}_{-1269}$ & 	& -0.171	 $\pm	0.016	$  & 	-0.051	 $\pm	0.024	$  & 	0.179	 $\pm	0.029	$  & 	$0.69\pm$ 0.06  & $5122^{+837}_{-831}$		\\ 
C2\tablenotemark{*} & 	262.67685	 & 	-21.497011	 & 0.41 & $	-175^{+672}_{-700}$ & 	& -0.047	 $\pm	0.016	$  & 	-0.013	 $\pm	0.024	$  & 	0.049	 $\pm	0.028	$  & 	$0.83\pm$ 0.30  & $1406^{+795}_{-796}$		\\
N7	 & 	262.65918	 & 	-21.466017	 & 1.71 & $-225^{+382}_{-398}$ & $	244	_{	-10	}^{	+46    	}$	& 0.026	 $\pm	0.015	$  & 	0.026	 $\pm	0.023	$  & 	0.037	 $\pm	0.028	$  & 	$0.15\pm $0.01  & $1077^{+783}_{-784}$	\\ 
C4\tablenotemark{*}	 & 	262.66782	 & 	-21.489190	 & 0.29 & $	-233^{+803}_{-862}$ & 	& 0.004	 $\pm	0.015	$  & 	0.004	 $\pm	0.023	$  & 	0.006	 $\pm	0.028	$  & 	$0.14\pm$ 0.07  & $	289^{+801}_{-840}$		\\ 
S1	 & 	262.66758	 & 	-21.517109	 & 1.54 & $-246^{+897}_{-882}$ & 	& -0.033	 $\pm	0.018	$  & 	-0.070	 $\pm	0.024	$  & 	0.077	 $\pm	0.03	$  & 	$0.34\pm$ 0.03  & $2205^{+854}_{-854}$		\\ 
S2	 & 	262.66207	 & 	-21.512892	 & 1.38 & $	-536^{+1060}_{-1067}$ & 	& 0.058	 $\pm	0.017	$  & 	-0.120	 $\pm	0.024	$  & 	0.133	 $\pm	0.029	$  & 	$0.66\pm$ 0.07  & $	3823^{+831}_{-831}$		\\ 
Ear3	 & 	262.63949	 & 	-21.488767	 & 1.83 & $-2239^{+1087}_{-973}$ & 	& 0.140	 $\pm	0.016	$  & 	0.018	 $\pm	0.024	$  & 	0.141	 $\pm	0.03	$  & 	$0.53\pm$ 0.04  & $4595^{+915}_{-884}$		\\ 
N5	 & 	262.68243	 & 	-21.475636	 &  1.13 &$-6716^{+1535}_{-1613}$ & 	& -0.054	 $\pm	0.018	$  & 	0.077	 $\pm	0.024	$  & 	0.094	 $\pm	0.03	$  & 	$0.57\pm$ 0.07  & $7229^{+1461}_{-1531}$		\\
\enddata
\tablenotetext{*}{CSM-dominated knot.}
%\tablenotetext{\dagger}{Knot labels in parentheses are those used by \citet{2017ApJ...845..167S}.}
\tablenotetext{a}{Position in 2016 (J2000).}
\tablenotetext{b}{Projected angular distance from kinematic center estimated by \citet{2017ApJ...845..167S}; R.A.(J2000) = 17\textsuperscript{h} 30\textsuperscript{m} 41\textsuperscript{s}.321 and Declination(J2000) = -21$^{\circ}$ 29\arcmin{} 30\arcsec.51, with uncertainties of $\sigma_{R.A.}= \pm$  0.073\arcmin{} and $\sigma_{Dec}= \pm$ 0.072\arcmin{}, respectively. }
\tablenotetext{c}{Values taken from \citet{2017ApJ...845..167S}. Errors represent a 68\%{} confidence interval.}
\tablenotetext{d}{$\mu_{Tot} = \sqrt[]{\mu_{R.A.}^{2}+\mu_{Dec}^{2}}$.}
\tablenotetext{e}{Expansion index (see Section \ref{subsec:propmot}).}
\tablenotetext{f}{Estimated space velocity for a distance of 6 kpc.}
\label{table:all}
\end{deluxetable*}

Given that Si and Fe are the most efficiently produced elements in a Type Ia explosion, we identified our knots as CSM-dominated or ejecta-dominated based on our measured abundances of Si and Fe. Knots with low Si and Fe abundances relative to solar values, [Si/Si\textsubscript{\(\odot\)}] $\lesssim$ 1.5 and [Fe/Fe\textsubscript{\(\odot\)}] $\lesssim$ 1, were deemed CSM-dominated, while those which have an enhanced abundance [Si/Si\textsubscript{\(\odot\)}] $\gtrsim$ 3, and [Fe/Fe\textsubscript{\(\odot\)}] $>$ 1, were classified to be ejecta-dominated. This way, we identified 15 knots as ejecta-dominated and two knots as CSM-dominated. The best-fit electron temperatures of nearly all ejecta knots in our sample are {\it kT} $\sim$ 2 - 5 keV, with ionization timescales {\it n\textsubscript{e}t} $\sim$ 1 - 3 $\times$ 10\textsuperscript{10} cm\textsuperscript{-3} s. The medians of these best-fit {\it kT} and {\it n\textsubscript{e}t} ejecta values generally agree with the higher-temperature ejecta components measured by \citet{2015ApJ...808...49K}. For three ejecta-dominated knots, S2, N7, and Ear1, and for the CSM-dominated knots, we measure lower temperatures ({\it kT} $\sim$ 0.5 - 1.3 keV) and higher ionization timescales ({\it n\textsubscript{e}t} $\sim$ 5 $\times$ 10\textsuperscript{10} - 10\textsuperscript{12} cm\textsuperscript{-3} s). We attribute the outlying {\it kT} and {\it n\textsubscript{e}t} observed in these three ejecta knots to possible CSM interaction. The spectral fitting results are summarized in Table \ref{table:spec_table} in the Appendix.

%% The "ht!" tells LaTeX to put the figure "here" first, at the "top" next
%% and to override the normal way of calculating a float position

\subsection{Proper Motions} \label{subsec:propmot}

\begin{figure*}[!htbp]
\gridline{\rightfig{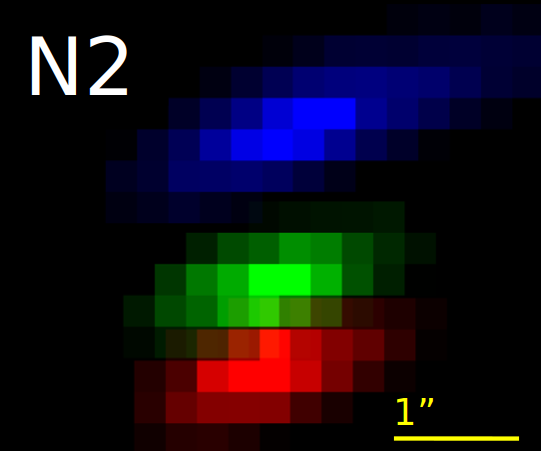}{0.25\textwidth}{(a)}
          \leftfig{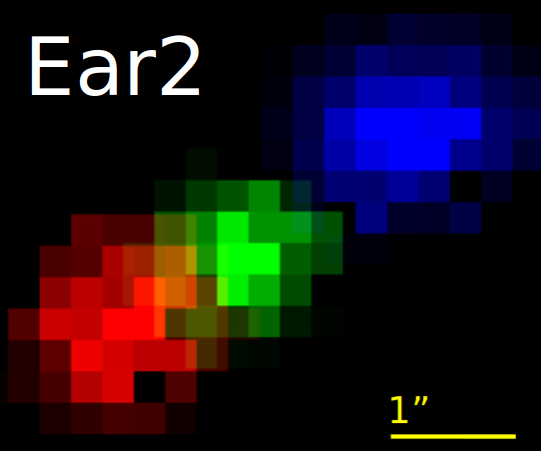}{0.25\textwidth}{(b)}
		 }
\gridline{\rightfig{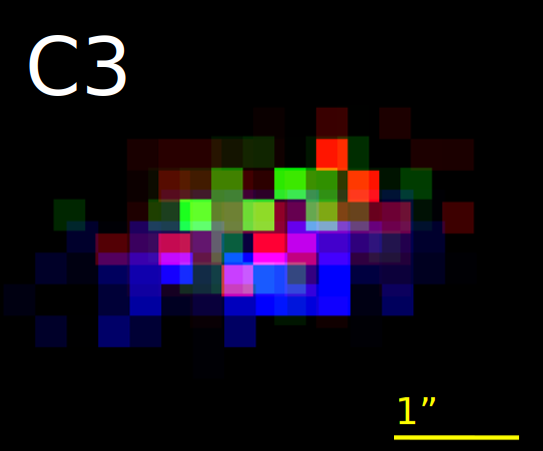}{0.25\textwidth}{(c)}
          \leftfig{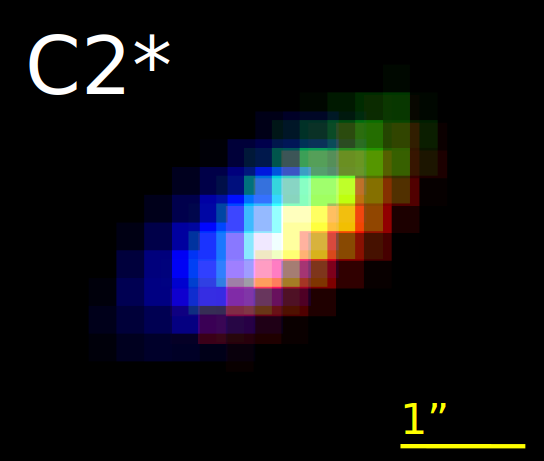}{0.246\textwidth}{(d)}
         }
\caption{$\Delta$C (where $\Delta$C is the difference between the C statistic for this image and the minimum C statistic, as defined in Section \ref{subsec:propmot}) images showing positional differences of regions (a) N2, (b) Ear2, (c) C3, and (d) C2* among observations performed in 2000 (red), 2004 (green), and 2014 (blue). \label{fig:deltac}}   
\end{figure*}

Based on the archival {\it Chandra} ACIS data covering the net time-span of 14 years (2000-2014, Table \ref{table:obs}), we estimate the proper motions of the small ejecta regions for which we measure {\it v\textsubscript{r}}. To measure the proper motions, we apply the methods used in \citet{2018ApJ...853...46S}. We took the image from the long observation in 2006 as the reference ``model" for each knot, compared it to the images from other epochs by incrementally shifting it in R.A. and declination, and then calculated the value of the Cash statistic \citep{1979ApJ...228..939C},
\begin{equation} \label{eq:cstat}
C = -2\sum_{i,j}(n_{i,j}  \ln m_{i,j} - m_{i,j} - \ln n_{i,j}!),
\end{equation}
where $n_{i,j}$ and $m_{i,j}$ are the number of counts in the $i^{th}$, $j^{th}$ pixel from the current epoch, and in 2006, respectively, scaled by the total number of counts in the SNR.  When the Cash statistic reached a minimum value, it means the pixel values in the image for each ``test" epoch most closely matched those found in the reference image, indicating that its position in the test epoch was determined. We estimate the error in the parameters using $\Delta C = C - C_{min}$, which may be interpreted in a way similar to $\Delta \chi^{2}$, and also include the systematic image alignment uncertainty from each epoch \citep{2017ApJ...845..167S}. The results of our proper motion measurements are summarized in Table \ref{table:all}. Our measured proper motions range $\mu_{R.A.} \sim$ -0.17\arcsec{} yr\textsuperscript{-1} to 0.17\arcsec{} yr\textsuperscript{-1} in R.A.\ and $\mu_{Dec} \sim$ -0.12\arcsec{} yr\textsuperscript{-1} to 0.14\arcsec{} yr\textsuperscript{-1} in declination. Figure \ref{fig:deltac} shows zoom-in images of knots, demonstrating their positional changes over 14 years. Knot Ear2, projected within the western Ear of Kepler shows the largest proper motion,  $ \mu_{Tot} \sim$ 0.20\arcsec{} yr\textsuperscript{-1}, which is perhaps as expected, considering that it is an ejecta knot projected beyond the main shell of Kepler.  The CSM-dominated regions generally show negligible proper motions, which may also be expected.

\citet{2017ApJ...845..167S} found that ejecta knots with the highest {\it v\textsubscript{r}} (N2, N1, N3) tend to show proper motions close to their extrapolated time-averaged rates for the change of angular positions, $\mu_{Avg}$ (their angular distance from the SNR center estimated by \citet{2017ApJ...845..167S} divided by the age of Kepler, 412 years as of 2016), suggesting that they have not undergone significant deceleration since the explosion (i.e. they are nearly freely expanding). From here on we refer to $\mu_{Tot}/\mu_{Avg}$ as the expansion index, $\eta$.  If an ejecta knot has been moving undecelerated since the explosion, we may expect $\eta$ $\approx$ 1. We find several ejecta knots to have an expansion index close to 1 ($\eta{} \gtrsim$ 0.7, see Table \ref{table:all}). We note that region C3 is an anomaly with $\eta$ = 1.86. This discrepancy is probably due to its projected proximity to the SNR center.  The angular offset of C3 from the SNR center is similar to the uncertainties on the SNR center position, and, in fact, $\eta$ is not constrained (Table \ref{table:all}). Knot C2* also shows a high $\eta$ value, and is projected near the center of the SNR with a large uncertainty in $\eta$ ($\pm{} 0.3$). Its spectrum is clearly CSM-dominated and its low proper motion is consistent with a CSM origin. In general, CSM-dominated regions are not expected to have a high $\eta$ value. The source of this discrepancy is unclear, however, we speculate that this dense filament of CSM-dominated gas may have been ejected from the progenitor system shortly before the SN explosion took place. Thus, like other parts of the remnant, it has possibly only been traveling for $\sim{}$ 400 years.

\section{Discussion} \label{sec:discuss}

\subsection{Distance to Kepler} \label{subsec:distance}

The kinematic nature of ejecta knot N2 in Kepler is remarkable. \citet{2017ApJ...845..167S} measured an expansion index $\eta \sim 1$, indicating that it is almost freely-expanding.  Here, we measure a similarly high expansion index, and a high {\it v\textsubscript{r}} (nearly 8,000  km s\textsuperscript{-1}). In general, X-ray-emitting ejecta features in SNRs are expected to be heated to $T >$ 10\textsuperscript{6} K by the reverse shock, being somewhat decelerated in the process. To explain the existence of nearly freely-expanding ejecta knots in Kepler, \citet{2017ApJ...845..167S} used the findings from \citet{2001ApJ...549.1119W} to argue that these ejecta knots may have survived to the current age of Kepler ($\sim$ 400 years) if their initial density contrast to the surrounding medium was high ($\gtrsim$ 100). Alternatively, they suggested that the highly-structured environment of the remnant contains low density ($n_{H} \sim$ 0.1 cm\textsuperscript{-3}) ``windows'' through which some ejecta knots may have traveled. This causes a late encounter with the reverse shock, allowing for the survivability of lower density-contrast, nearly undecelerated knots to the forward shock region, according to the simulations of \citet{2001ApJ...549.1119W}.  This is the scenario favored by \citet{2017ApJ...845..167S}. Either scenario may be applied in the interpretation of our results. Considering that Kepler is located hundreds of parsecs out of the Galactic plane where the ambient density is $n_{H} \lesssim$ 0.01 cm\textsuperscript{-3} \citep{1977ApJ...218..148M}, the existence of low density regions around the SN site appears to be plausible.   

Since we know the exact age of the SNR, and we have measured the radial velocity and the projected angular distance from the center of the remnant, only the inclination angle of the nearly freely-moving knot's velocity vector against the line of sight needs to be constrained in order
to estimate the distance. There are ejecta-dominated regions projected close to the outermost boundary of the main shell (e.g. N7) and even beyond it (Ear2). These knots show smaller expansion indices (i.e. stronger deceleration) than N2, while N2 has an apparent nearly constant proper motion since the explosion, and unusually high {\it v\textsubscript{r}}. The forward shock itself has significantly decelerated: \citet{2008ApJ...689..231V} and \citet{2008ApJ...689..225K} found an average expansion parameter of $\sim$ 0.5 to 0.6. Hence, it seems likely that N2 may have reached near or even beyond the main shell, similar to ejecta ``bullets'' reported in other SNRs \citep[e.g.,][]{1995Natur.373..590S, 2012ApJ...748..117P, 2014ApJ...781...65W}.

 Depending on the location of N2, the inclination angle for its space velocity vector against the line of sight may be constrained. We assume three cases for the physical location of N2:  1) at the outermost boundary of the SNR's main shell (the projected angular distance from the SNR center $D \sim{} 1.8\arcmin{}$), 2) at the physical distance corresponding to the angular distance (from the SNR center) to the western Ear's outermost boundary, i.e., the  visible maximum angular extent of the X-ray emission ($D \sim{} 2.3\arcmin{}$), and 3) a location significantly beyond the main SNR shell at the distance corresponding to $\sim$ 1.5 times the radius of the SNR's main shell ($D \sim{} 2.7\arcmin{}$).  The expansion center of Kepler's SNR has been estimated in radio wavelengths by \citet{1984ApJ...287..295M} and \citet{2002ApJ...580..914D}, and later in X-rays by \citet{2008ApJ...689..225K} and  \citet{2008ApJ...689..231V}. Recently, \citet{2017ApJ...845..167S} estimated two possible expansion centers by tracing back the proper motion of a few ejecta knots with high expansion indices to a common origin, one assuming no deceleration, and the other a power-law evolution of radius with time (i.e., deceleration). We take the ``decelerated'' kinematic center estimated by   \citet{2017ApJ...845..167S} as the explosion site unless otherwise noted. We may calculate the distance to Kepler by considering that N2 has been moving with our measured {\it v\textsubscript{r}} along the line of sight since the explosion. This approach would almost certainly result in an underestimate of the true distance, however since we are in general more interested in a lower limit to the distance, this assumption would not affect our conclusions.

In  scenarios 1, 2, and 3,  we assume that N2 has reached the main shell of the remnant, or beyond, and thus we estimate distances of  $d$ $\sim$ 7.5 kpc, $\sim$ 5.4 kpc, and $\sim$ 4.4 kpc, respectively.  Recently, \citet{2017ApJ...842..112R} interpreted historical light curves of SN 1604, and \citet{2016ApJ...817...36S} used proper motion and line width measurements of Balmer filaments to independently estimate a distance range, $ d \sim$ 5.1 $\pm$ 0.8 kpc to Kepler. Our distance range is generally consistent with this value, and also with somewhat farther distance estimates which suggest an energetic Type Ia explosion for SN 1604 \citep{2008A&A...488..219A,2008ApJ...689..231V,2012A&A...537A.139C,2012ApJ...756....6P,2015ApJ...808...49K}. Considering the amount of \textsuperscript{56}Ni required to explain the bulk properties of the X-ray spectrum, the spectral and hydrodynamical fitting done by \citet{2012ApJ...756....6P} and \citet{2015ApJ...808...49K} suggests that the data are incompatible with a normal Type Ia explosion, but may be consistent with a DDTa model, which is more energetic. Since the age is known, this places the SNR at a distance of $>$ 5 kpc. However, \citet{2017ApJ...842..112R} argues that the best-fit stretch factor to the historical light curve indicates that it is more consistent with a normal Type Ia SN. Considering that our estimated lower limit (scenario 3) is likely to be conservative, and unless we have a relatively unique viewing angle, it is reasonable that knot N2 is located nearby, or less than, a distance from the center of the remnant described in scenario 2. Thus, we may conclude that our $v_{r}$ measurement suggests $d \gtrsim$ 5 kpc, and hence tends to favor the distance estimates which suggest an energetic Type Ia explosion for SN 1604. 

Although it may not be favored due to our measured high proper motion and {\it v\textsubscript{r}}, for completeness, we may consider that even the fastest ejecta knots (e.g., N2) in Kepler have been significantly decelerated rather than nearly freely expanding. In this scenario, the ejecta knot is heated between the forward and reverse shocks as expected by standard SNR dynamics \citep{1982ApJ...259L..85C}, and it would be traveling generally with the bulk of ejecta gas in the SNR. With this configuration, a longer distance to Kepler is implied ($d \sim$ 11.0 kpc), which we may consider to be a conservative upper limit.

\subsection{Velocity Distribution of Ejecta} \label{subsec:vel_distribution}

Based on our {\it v\textsubscript{r}} and proper motion measurements, we measure space velocities, {\it v\textsubscript{s}} $\sim$ (1,100 - 8,700){\it d}\textsubscript{6} km s\textsuperscript{-1} (with {\it d}\textsubscript{6} in units of 6 kpc), with an average velocity, {\it v\textsubscript{s}} $\sim$ 4,600{\it d}\textsubscript{6} km s\textsuperscript{-1}, for the 15 individual ejecta knots.  The fastest known stars in the Milky Way  (which are probably ejected from SN explosions in white dwarf binaries) show space velocities of $\sim$ 2,000 km s\textsuperscript{-1} \citep{2018ApJ...865...15S}. Thus, the velocities we obtain for several knots are highly significant, and cannot be attributed to a systemic velocity for the SNR.

The broad range of ejecta space velocities and expansion indices (see Table \ref{table:obs}) that we measure in our sample may be characteristic for an SNR transitioning from the free-expansion to Sedov-Taylor phase.  Measurements of the proper motion at various locations along Kepler's forward shock by \citet{2008ApJ...689..225K} and  \citet{2008ApJ...689..231V} found expansion indices of 0.47 - 0.82 and 0.3 - 0.7, respectively. For remnants nearing the Sedov-Taylor phase,  \citet{1982ApJ...259L..85C} estimated $\eta$ = 0.4 for $s$ = 0 and $\eta$ = 0.67 for $s$ = 2 ambient density power-law solutions.  Hence, our ejecta velocity measurements and previous forward shock analyses apparently suggest that the kinematics for some regions in Kepler may be dominated by nearly free-expansion, while others are better described by Sedov-Taylor dynamics. New 3-D hydrodynamical simulations that focus on Kepler and these high $\eta$ knots may give some insight into their origin.

    The knots N2, N5, N1, and N3  have the highest measured space velocities (6,100 - 8,700{\it d}\textsubscript{6} km s\textsuperscript{-1}), and are all located in the ``steep arc" \citep{2002ApJ...580..914D} of Kepler's SNR, a ``bar'' of bright X-ray emission which runs from east to west, located about halfway between the center of the remnant and the outer edge of the main shell. They are projected close to each other within a small ($50\arcsec{}\times{} 20\arcsec$) area. This proximity, and similarities in their measured Si abundances, space velocity vectors, and expansion indices, suggest that these knots might have originated generally from a ``common'' layer of the exploding white dwarf. \citet{2017ApJ...845..167S} measured properties of another knot (they label ``N4'') projected within the steep arc, which exhibited similar properties to N2, N5, N1, and N3. This suggests that ejecta within the steep arc have generally homogeneous kinematic and spectroscopic properties.

In the western Ear, we measure high space velocity in region Ear2, {\it v\textsubscript{s}} $\sim$ 5,800{\it d}\textsubscript{6} km s\textsuperscript{-1}. Such a high velocity may be expected considering the knot's projected location in the western Ear feature which protrudes out about 30\% beyond the main shell. Interestingly, knot Ear1 has a significantly smaller space velocity, {\it v\textsubscript{s}} $\sim$ 3,300{\it d}\textsubscript{6} km s\textsuperscript{-1} , even though it is projected very close to the position of Ear2. Knot Ear1 may be interacting with a CSM-dominated feature identified by \citet{2013ApJ...764...63B} projected adjacent to it, which could have caused it to significantly decelerate recently. Such an interaction between ejecta and CSM may produce H$\alpha$ emission. We searched for H$\alpha$ emission at the location of Ear1 in the archival {\it Hubble Space Telescope} images (with the F656N filter) of Kepler \citep{2016ApJ...817...36S}.  We found a faint wisp centered at Ear1's position, possibly indicating the presence of shocked CSM gas, which would support our conclusion of an ejecta-CSM interaction there.  

Considering their spatial proximity and similarly high Si abundance, it seems likely that Ear1 and Ear2 were produced very near to each other during the SN. It is interesting that these knots are projected $\sim{} 1\arcmin{}$ in decl. north of the center of the remnant, as are the ejecta knots in the steep arc. In our distance estimation, we assumed that knot N2 is located at or beyond the main shell. Thus, if we viewed Kepler at a different angle, it may appear as though the steep arc and western Ear are similar structures extending to different directions. This morphological interpretation may not be consistent with the bipolar-outflow scenario \citep{2013MNRAS.435..320T} as the origin of the Ears. However, we measured generally higher Si abundances in the western Ear than in the steep arc (roughly by a factor of $\sim 5$), as did \citet{2019ApJ...872...45S}, who recently reported a similar result. This abundance discrepancy is not in line with the scenario that the Ear and arc features share a common physical origin. Thus, while we find intriguing similarities in kinematic properties between these substructures of Kepler, their true physical origins remain unanswered. Detailed hydrodynamic simulations may be needed to test these scenarios, which are beyond the scope of this work.

\begin{figure*}[!htbp]
%\gridline{\rightfig{vr_vs_r_plot_ref2.eps}{0.3\textwidth}{(a)}
%          \leftfig{side_Kepler_04_19_2019_withLabels.png}{0.43\textwidth}{(b)}
%        }
\begin{tabular}{cc}
  \includegraphics[height=0.41\textheight]{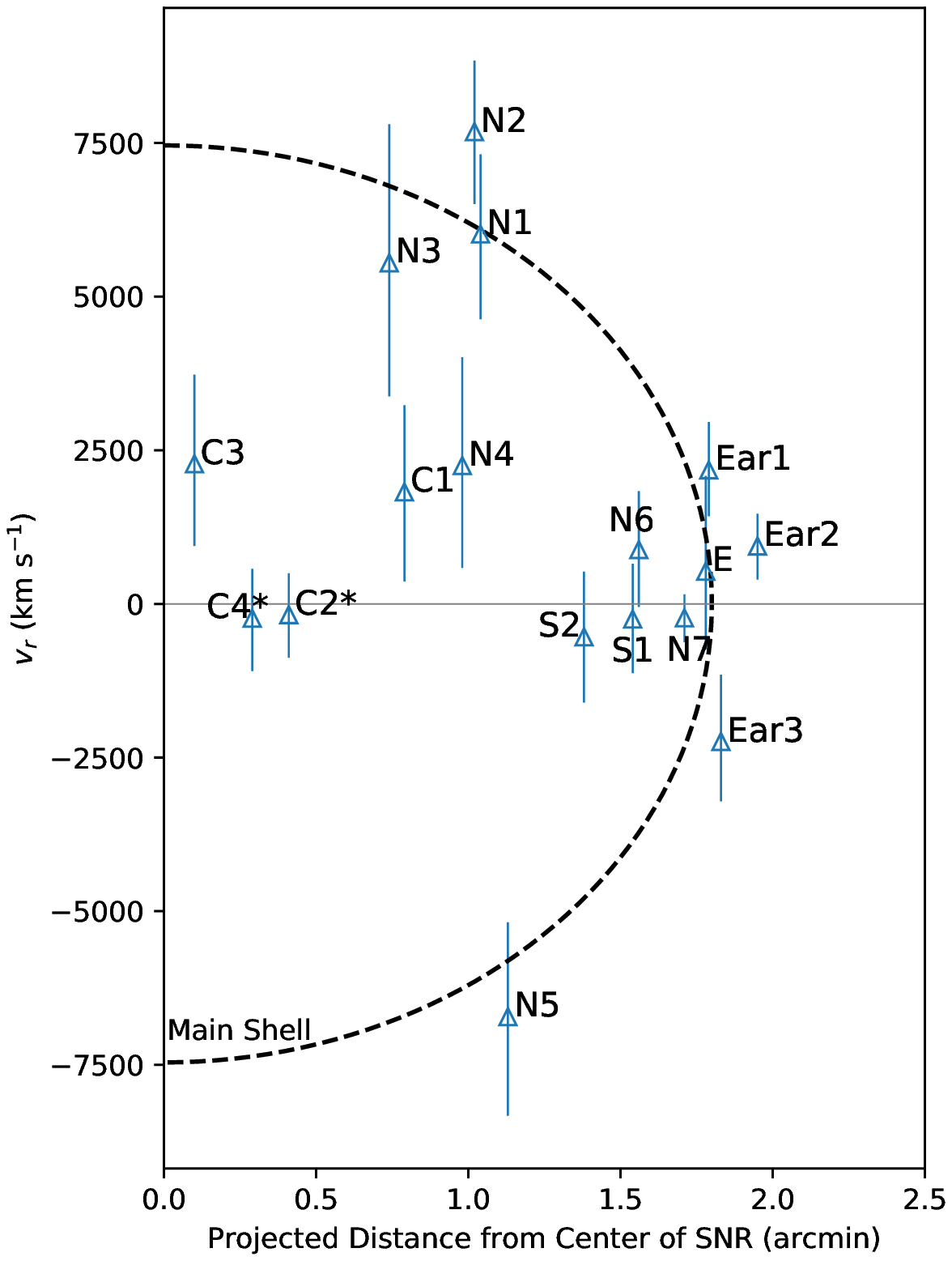}
  &
  \includegraphics[height=0.38\textheight]{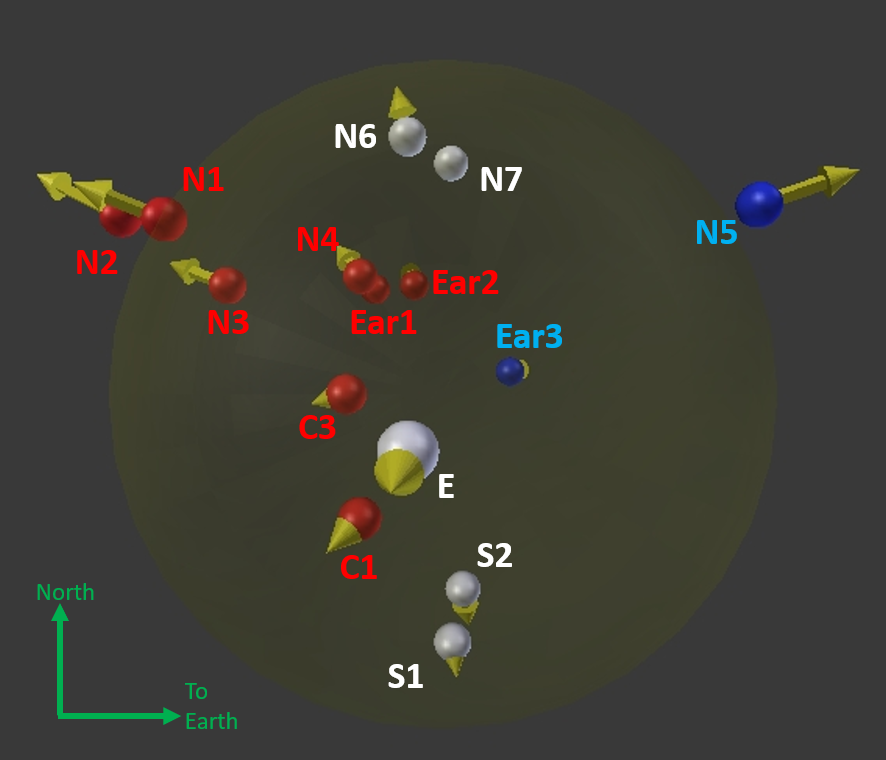}
   \\                                                     
   (a) & (b)
\end{tabular}
\caption{Panel (a) shows the positions of ejecta knots in {\it v\textsubscript{r}} vs. {\it r} (projected angular distance from the center of the SNR) space. The dashed line is the approximate location of the outermost boundary of the main SNR shell. Panel (b) shows a 3-D perspective of the locations of our measured ejecta knots. The red spheres represent redshifted knots, blue spheres are blueshifted knots, and white spheres are those with negligible Doppler shift.  The gold arrows indicate the knots' relative magnitude of space velocites and directions. The shaded circle shows the approximate location of the main shell of Kepler's SNR.\label{fig:asymm}}   
\end{figure*}

The HETG spectra of ten ejecta regions from our sample show a significant Doppler shift (i.e., $|{\it v\textsubscript{r}}| \gtrsim{} 10^{3}$ km s\textsuperscript{-1}). The majority of them (eight regions) are redshifted.  This may suggest a significantly asymmetric velocity distribution of ejecta knots along the line of sight (see Figure \ref{fig:asymm}). However, we note that our sample size of the ejecta knots is limited. In particular, our sample regions offer very little coverage in the southern shell of the SNR. Thus, the apparent asymmetric ejecta distribution along the line of sight might have been a selection effect. To make a conclusive statement regarding the overall 3-D distribution of Si-rich ejecta in Kepler, a significantly larger sample of high-resolution velocity measurements from ejecta regions across the entire face of the SNR is required.  A significantly deeper {\it Chandra} HETG observation would be needed to achieve this. Nonetheless, it is interesting to note that some similar uneven ejecta distributions have been reported in studies using the lower-resolution X-ray CCD spectroscopy from archival {\it Chandra} ACIS data. \citet{2017ApJ...845..167S} found that only two ejecta-dominated knots out of the eleven (four in common with this work) included in their study were significantly blue-shifted. Those regions show relatively weak He-like Si K$\alpha$ emission line fluxes, and thus, we could not measure their {\it v\textsubscript{r}} using our HETGS data  due to low photon count statistics.  \citet{2018PASJ...70...88K} reported that a general asymmetry exists in the Fe-rich ejecta along the line of sight in Kepler. \citet{2013ApJ...764...63B} suggested that the asymmetry in Fe ejecta across the face of the SNR could be a result of ejecta being blocked by the progenitor's companion star. While our results suggesting an uneven line-of-sight ejecta distribution cannot be conclusive based on the current data, previous studies of Kepler appear to be consistent with our results.  

The suggested asymmetric distribution of the ejecta (if it is confirmed) could be the result of Kepler's interactions with its nonuniform surroundings.  \citet{2012ApJ...756....6P} and \citet{2007ApJ...662..998B} argued that the north-south density gradient they found in the surrounding medium of Kepler is required to explain the observed bowshock in the north of the remnant, and the infrared intensity variation between the northern and southern rims. Such a density gradient across the near and far sides of the remnant, with surrounding material on the near side having a lower density on average, could lead to an under-developed or late reverse shock, causing blueshifted knots to appear fainter. Alternatively, the tentative asymmetric ejecta distribution along the line of sight might have been caused by a true asymmetry in the SN explosion itself. The global asymmetry in Type Ia SNe may in general be caused by the strength and geometry of ignition of the SN explosion \citep{2010Natur.466...82M}.  The validity and true nature of the asymmetric ejecta distribution in Kepler's SNR that we observe in the Chandra data are unclear due to our small sample size. Follow-up {\it Chandra} HETGS observations of Kepler with deeper exposures would be warranted to perform a more extensive census of the ejecta velocity distribution (significantly beyond the capacity of the existing ACIS and HETG data) throughout the entire SNR, which is required to reveal the true 3-D nature of Kepler's SN explosion.

\section{Conclusions} \label{sec:conclusions}

We have measured the radial velocities and proper motions of 17 small emission features (15 ejecta and 2 CSM knots) in Kepler's supernova remnant using our {\it Chandra} HETGS observation and the archival {\it Chandra} ACIS data.  We find that a handful of knots are moving at speeds approaching $\sim{} 10^{4}$ km s\textsuperscript{-1}, with expansion indices approaching $\eta{} \sim$ 1, indicating nearly a free expansion.  Based on our radial velocity measurement of such a fast-moving ejecta knot, we estimate the distance to Kepler. While our distance estimates may vary depending on our assumption of the degree of deceleration of the ejecta knot ($d \sim$ 4.4 - 7.0 kpc), a relatively long distance of $d >$ 5 kpc is favored. Our estimated distance range generally supports an energetic Type Ia SN for Kepler.  

We note that most of our {\it v\textsubscript{r}} measurements indicate a redshifted spectrum, suggesting an asymmetry in the along-the-line-of-sight ejecta distribution of the remnant. However, this study involves only a small sample of ejecta knots, most of which are projected in the northern shell of the SNR. Thus, while it provides hints into some intriguing kinematic characteristics of the Type Ia SN explosion which created Kepler, this work is limited in revealing the true 3-D structure of the entire SNR. A longer observation of Kepler using the {\it Chandra} HETGS would be required to measure {\it v\textsubscript{r}} for a significantly larger number of ejecta knots covering the entire face of the SNR. Such measurements would yield a detailed picture of the 3-D distribution of ejecta, and provide observational constraints for more realistic Type Ia SN models.

%Thus, the utility of our study, while revealing a hint of some intriguing kinematic characteristics of the Type Ia SN explosion which created Kepler, is limited in revealing the true 3-D structure of the entire SNR.
%Our study provides hints into the intriguing kinematic characteristics of the Type Ia SN explosion which created Kepler, and is a valuable first step towards revealing the true 3-D structure of the entire SNR.

%which would be be essential to provide observational constraints to establish more realistic Type Ia SN models

\bigskip
This work has been supported in part by NASA {\it Chandra} Grants GO6-17060X and AR7-18006X.  J.P.H. acknowledges support for supernova remnant research from NASA grant NNX15AK71G to Rutgers University. T.S. was supported by the Japan Society for the Promotion of Science (JSPS) KAKENHI Grant Number JP19K14739 and the Special Postdoctoral Researchers Program in RIKEN. We also thank the anonymous referees for providing valuable input which strengthened this paper.

\bigskip
\bigskip
\bigskip
\bigskip




\counterwithin{figure}{section}
\section*{Appendix}

%\appendix
\counterwithin{figure}{section}
%\numberwithin{table}{section}

\setcounter{figure}{0}
\renewcommand{\thefigure}{A\arabic{figure}}

\begin{figure}[!htpb]
\includegraphics[height=0.6\textwidth]{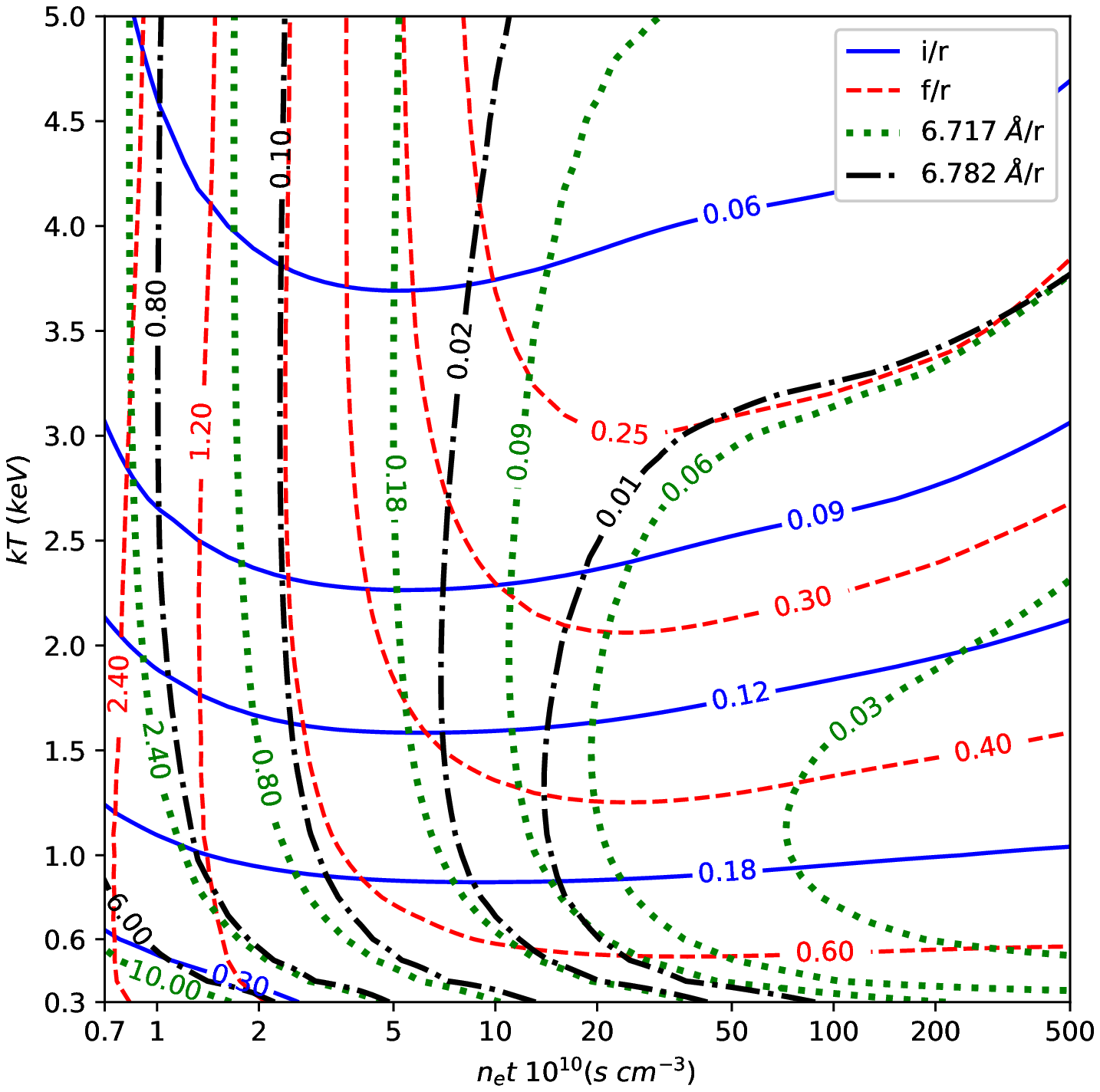}
\centering
\caption{Contours of line emission ratios for various temperatures and ionization timescales.  The blue solid and red dashed contours represent ratio values of of Si XIII intercombination to resonance, and forbidden to resonance line fluxes, respectively.  The green dotted and black dashed-dotted contours show ratio values of Si XII emission lines 6.717 $\AA$ and 6.782 $\AA$, to the Si XIII resonance line ({\it r}), respectively. }
\label{fig:ratio_cont}
\end{figure}

\subsection*{Line Emission Ratios}

To determine the line emission contributions from electronic shell transitions of Si XIII and Si XII, we measured the temperature and ionization timescale of each knot by fitting an absorbed plane shock model to its broadband (0.3 - 7.0 keV) ACIS spectrum (see Section \ref{subsec:specmod}). We show our results in Table \ref{table:spec_table}. Based on each knot's best-fit temperature and ionization timescale, we used a {\tt phabs*vvpshock} model to calculate the line ratios using XSPEC (version 12.10) with  NEI APEC spectral data (version 3.0.9) \citep{1996ASPC..101...17A,2001ApJ...556L..91S}. Our best-fit ACIS broadband models (see Section \ref{subsec:specmod}) suggest ratio values ranging from $\sim$ 0.05 - 0.23 and 0.47 - 2.15 for {\it i/r} and {\it f/r}, respectively. The 6.717 \AA{} and 6.782 \AA{} Si XII line ratios vary from 0.03 - 1.96 and 8e-4 - 0.98, respectively. Figure A1 shows contour plots of each line ratio for a range of temperatures and ionization timescales. Varying the line ratio values used in each measurement does not significantly affect our main results for the high $v_{r}$ knots (N1 - N3, N5), and although our sample size is limited, the overall dominance of the redshift in the ejecta $v_{r}$ also does not change. Hence, the overall results of our HETG $v_{r}$ measurements are generally independent of the line ratio values estimated from the ACIS model fits.

\bigskip
\bigskip
\bigskip
\bigskip
\bigskip
\bigskip
\bigskip
\bigskip
\bigskip

%\bigskip
%\bigskip
%\bigskip
%\bigskip
%We estimate the uncertainty from emission line ratio estimates to be a few hundred km s\textsuperscript{-1} on average, and thus does not affect our conclusions.

%\begin{figure}[htpb]
%\includegraphics[height=0.45\textwidth]{knot_ratio_diff_vr_01-04-2020.eps}
%\centering
%\caption{ The estimated radial velocities of each knot using measured Si XIII and Si XII emission line ratios which produce the greatest redshift (red cross), the greatest blueshift (blue circle), and each knot's measured best-fit values (green diamond). }
%\label{fig:vr_range}
%\end{figure}

%\subsection*{A1. ACIS Spectral Fitting Results}
% Start table numbering over, calling them A..
\setcounter{table}{0}
\renewcommand{\thetable}{A\arabic{table}}

\startlongtable
\begin{deluxetable*}{ccCccccc} 
\tablecaption{ACIS Spectral Fitting Results \label{tab:mathmode}}
\tablecolumns{12}
%\small
%\tablename{A.1}
%\tablenum{1}
\tablewidth{0pt}
\tabletypesize{\small}
\tablehead{
\colhead{Region} &
\colhead{{\it kT} (keV)} &
\colhead{$\tau$  ($10^{10}$s cm\textsuperscript{-3})\tablenotemark{a}} &
\colhead{Redshift ($10^{-2}$)} & 
\colhead{$K$ ($10^{8}$cm\textsuperscript{-5})\tablenotemark{b}} &
\colhead{$\chi^2/dof$}  &
\colhead{$i$/$r$\tablenotemark{c}} &
\colhead{$f$/$r$\tablenotemark{d}} %&
%\colhead{{\it 6.717/r}} & 
%\colhead{{\it 6.782/r}} 
}
\startdata
N2  &  $3.86^{+0.94}_{-0.75}$  &  $2.18^{+0.28}_{-0.23}$  &  $2.66^{+0.07}_{-0.22}$  &  $12.8^{+2.2}_{-1.9}$  &  191.2/138  &  0.06   & 0.67   \\%& 0.53 & 0.11 \\  
N1  & $2.15^{+0.41}_{-0.33}$  &    $2.32^{+0.30}_{-0.26}$  &  $2.47^{+0.01}_{-0.28}$  &  $9.99^{+2.06}_{-1.46}$  &  131.9/103  &  0.10  & 0.64  \\ %&  0.54 & 0.10 \\ 
N3  &  $3.49^{+0.69}_{-0.59}$  &   $2.16^{+0.17}_{-0.04}$  &  $2.13^{+0.07}_{-0.08}$  &  $10.4^{+1.2}_{-1.0}$  &  135.8/113  & 0.06   & 0.67   \\%& 0.54  & 0.11 \\  
C3  &  $2.44^{+0.49}_{-0.42}$  &   $0.99^{+0.22}_{-0.14}$ &  $0.15^{+0.05}_{-0.05}$  &  $8.27^{+1.04}_{-1.02}$  &  120.3/108  &  0.10  & 1.77   \\%& 1.94 & 0.86 \\ 
N4  &  $4.49^{+2.25}_{-0.72}$  &    $1.85^{+0.36}_{-0.29}$  &  $1.08^{+0.10}_{-0.01}$  &  $4.47^{+0.33}_{-0.25}$  &  144.9/104  &  0.05  &  0.85  \\%& 0.69 & 0.16 \\  
Ear1  &  $0.56^{+0.02}_{-0.04}$  &    $78.2^{+54.6}_{-19.2}$  &  $1.61^{+0.08}_{-0.05}$  & $5.78^{+1.14}_{-0.48}$  &  135.8/103  & 0.22   & 0.59  \\ %& 0.04  & 0.003 \\
C1  &  $3.64^{+1.03}_{-0.69}$  &    $2.11^{+0.28}_{-0.21}$ &  $0.31^{+0.04}_{-0.01}$  &  $4.76^{+0.64}_{-0.66}$  &  145.2/109  & 0.06   &  0.69  \\%& 0.56 & 0.12 \\
Ear2  &  $2.39^{+0.52}_{-0.38}$  &   $1.08^{+0.06}_{-0.12}$  &  $0.74^{+0.03}_{-0.04}$  &  $4.81^{+1.15}_{-0.57}$  &  114.1/112  & 0.10   &  1.58  \\%& 1.72 & 0.68 \\
N6  &  $3.34^{+0.88}_{-0.87}$  &   $2.58^{+0.52}_{-0.22}$  &  $0.80^{+0.01}_{-0.18}$  &  $3.07^{+0.40}_{-0.39}$  &  135.9/98  & 0.07   & 0.55   \\%& 0.42 & 0.08 \\ 
E  &  $3.05^{+0.39}_{-0.30}$  &   $1.52^{+0.27}_{-0.09}$  &  $1.06^{+0.01}_{-0.26}$  &  $29.3^{+1.6}_{-2.6}$  &  159.0/137  &  0.08  & 1.03  \\%&  0.96 &  0.26 \\
C2*  &  $1.27^{+0.12}_{-0.11}$  &   $5.15^{+1.34}_{-0.77}$  &  $0.19^{+1.10}_{-0.18}$  &  $29.8^{+2.8}_{-1.2}$  &  115.8/109  & 0.14   &  0.47  \\%& 0.22 & 0.03 \\ 
N7  &  $0.80^{+0.01}_{-0.02}$  &   $73.5^{+14.9}_{-11.2}$  &  $0.14^{+0.01}_{-0.17}$  &  $108^{+7}_{-5}$  &  194.9/143  & 0.19   & 0.51   \\%& 0.03  & 0.002 \\
C4*  &  $0.80^{+0.06}_{-0.07}$  &    $12.5^{+2.6}_{-2.4}$   &  $-0.33^{+0.02}_{-0.08}$  &  $97.1^{+8.4}_{-9.1}$  &  165.2/119  &  0.19  & 0.52   \\%& 0.11 & 0.01 \\
S1  &  $2.16^{+0.63}_{-0.59}$  &  $1.08^{+0.18}_{-0.12}$  &  $-0.49^{+0.12}_{-0.01}$  &  $1.38^{+0.53}_{-0.24}$  &  126.1/92  & 0.10   & 1.56   \\%& 1.77 & 0.70 \\ 
S2  &  $0.53^{+0.02}_{-0.02}$  &   $300^{+1590}_{-131}$  &  $-0.50^{+0.03}_{-0.04}$  &  $11.4^{+1.9}_{-0.3}$  &  137.0/107  &  0.23  &  0.61  \\%& 0.03 & 0.0008 \\ 
Ear3  &  $4.02^{+1.73}_{-1.04}$  &    $0.94^{+0.12}_{-0.09}$   &  $-1.18^{+0.05}_{-0.11}$  &  $3.36^{+0.35}_{-0.66}$  &  139.9/107  & 0.07   & 2.15   \\%& 1.96 & 0.98  \\ 
N5  &  $4.93^{+1.84}_{-1.32}$  &   $1.16^{+0.12}_{-0.13}$  &  $-1.68^{+0.06}_{-0.02}$  &  $4.02^{+1.04}_{-1.02}$  &  146.4/113  &  0.05  & 1.68   \\%& 1.43 & 0.56 \\
\enddata
\tablenotetext{a}{$\tau = n_et$, where $n_e$ is the electron density, and $t$ is the time since the plasma was shocked.}
\tablenotetext{b}{$K = \int n_{e}n_{H}dV/4\pi d^2$, where $n_{H}$ is the hydrogen density, $V$ is the volume of the region, and $d$ is the distance to the region.}
\tablenotetext{c}{Si K$\alpha{}$ intercombination ($i$) to resonance ($r$) line ratio.}
\tablenotetext{d}{Si K$\alpha{}$ forbidden ($f$) to resonance line ratio.}
\label{table:spec_table}
\end{deluxetable*}

\end{document}